# The Role of Community Building and Education as Key Pillar of Institutionalizing Responsible Quantum

Sanjay Vishwakarma*‡, Vishal Sharathchandra Bajpe *‡,
Ryan Mandelbaum*‡, Yuri Kobayashi*‡, Olivia Lanes*‡, and Mira Luca Wolf-Bauwens†‡

*IBM Quantum, Yorktown Heights, NY, USA
†IBM Research, Zurich, Rüschlikon, Switzerland
‡Authors contributed equally

*Abstract*—Quantum computing is an emerging technology whose positive and negative impacts on society are not yet fully known. As government, individuals, institutions, and corporations fund and develop this technology, they must ensure that they anticipate its impacts, prepare for its consequences, and steer its development in such a way that it enables the most good and prevents the most harm. However, individual stakeholders are not equipped to fully anticipate these consequences on their own — it requires a diverse community that is well-informed about quantum computing and its impacts. Collaborations and community-building across domains incorporating a variety of viewpoints, especially those from stakeholders most likely to be harmed, are fundamental pillars of developing and deploying quantum computing responsibly. This paper reviews responsible quantum computing proposals and literature, highlights the challenges in implementing these, and presents strategies developed at IBM aimed at building a diverse community of users and stakeholders to support the responsible development of this technology.

## I. Introduction

Quantum computing is a computer processing architecture that relies on the mathematics of quantum physics to tackle problems beyond the computational ability of today's supercomputers. Companies, research institutions, and governments are exploring the potential of quantum computing with the hope that it will bring value to society and the economy. A McKinsey report finds that the automotive, chemical, financial services, and life sciences industries could gain 1.35 trillion in value from quantum computing by 2035 [1].

Increased interest in quantum computing has led to concern that quantum computers may yield unforeseen negative impacts, or that positive impacts may lead to inequality between those with access to and funding for quantum, and those without [2].

Recently, institutions like the National Quantum Computing Centre in the UK (NQCC) and the World Economic Forum (WEF), and companies like IBM, Microsoft, and Google, have begun to perform research and publish governance models surrounding the responsible development and deployment of this new technology, a movement broadly called responsible quantum computing [3] [4] [5].

We understand responsible quantum computing as quantum computing that is aware of its effects. Responsible quantum computing practices include, for example, the publishing of principles to guide the development and deployment of quantum computing, incorporating lessons learned from other emerging technologies such as generative artificial intelligence, reviewing contracts and proposals to ensure that they carry clauses relating to responsible and acceptable uses of the technology, soliciting feedback from stakeholders, and spreading awareness of quantum computing and responsible quantum computing to those with less access to the technology [6].

A common principle incorporated into responsible quantum computing practices is considering the impacts of a diverse and inclusive set of stakeholders on the ways in which quantum computing is developed and deployed. Stakeholders are not limited to those with a direct interest in developing or using quantum computers, but include persons, entities, or groups of people that have indirect impact or are impacted by quantum computing. However, investment into quantum computing is spread unequally across the globe, with far more investment from countries in North America, Europe, and Asia versus those in South America and Africa [7]. Individuals from marginalized backgrounds in the United States continue to be underrepresented in the pool of physics degree recipients, and men far outnumber women as physics degree recipients. Therefore, voices from these underrepresented groups could be left out of conversations surrounding the potential impacts of quantum computing on stakeholders unless they are actively incorporated into responsible quantum computing practices. Finally, physics is widely considered a challenging topic to understand, and without active outreach and education efforts for the broader public, we risk leaving out those stakeholders without access to physics education [8].

In this paper, we review the proposals made in the literature with respect to incorporating more diverse voices into quantum computing and into responsible quantum computing practices. Then, we will present strategies incorporated by IBM to build a diverse, global community of quantum computing users and developers. This will include ongoing challenges in the way of building this community and possible solutions being pursued by our team. We will conclude with a summary of the work presented here and plans for IBM to continue building a diverse and inclusive pool of informed stakeholders to shape the future of quantum computing and responsible quantum computing.

## II. LITERATURE REVIEW

In the young literature on responsible quantum computing, core focus topics include education and diverse community building. These topics cover three main themes: avoiding the exacerbation of the digital divide, accessibility of open education, and education as a pillar on which a diverse workforce can be built. In this section of the paper, we will review some of the literature covering these three themes.

### A. Quantum computing should avoid exacerbating classical digital gaps

Learning from classical computing, a lack of focus on diverse community can widen the wider digital divide. Ten Holter et al (2022) [9] looks at how classical computing was deployed, analyses what we can learn from mistakes and missed opportunities for the case of quantum computing. Since quantum computing is likely to be a powerful tool, the paper urges that we ensure the benefits can be spread equally. It highlights initiatives meant to "bridge the gaps". These include free and open education programs, such as the Qubit by Qubit teaching program offered globally [9]. Thus, the paper demands that government-level efforts focus on equitable access to quantum computing and collaboration between governments and technology corporations to ensure open-access programs. It argues that where private companies profit from national education for their workforce, they should collaborate with national governments to ensure "democratization of such a key technology" [9, pp.4].

Similar demands are made in the report on the Multilateral Governance of Quantum Computing [10]. The report, penned by a group of policy makers reflecting on how quantum computing can be used to achieve the Sustainable Development Goals, highlights that existing national strategies for quantum technologies include a focus on education and workforce, but that more needs to be done in order to work against the "risk of a new digital divide" [10, pp.6]. In their findings, the authors emphasize that geopolitical developments have led to international collaboration no longer being encouraged "and that setting up diverse talent teams is increasingly difficult" [10, pp.6]. To mitigate these developments, the OQI report also calls for government action, the training of experts, and for initiatives that further draw attention to quantum technologies and their impact, such as the UNESCO International Year of Quantum 2025. It highlights that doing so would be part of more responsible innovation within the field of quantum.

### B. Quantum education should be widely accessible

While it is relatively easy to claim that quantum computing has been democratized, it has been argued that claims for democratization have been made too hastily [11]. In comparing existing efforts that focus on widening the access to quantum computing to definitions of democracy, Seskir et al. [11] argue that these efforts are required steps but are not sufficient. In particular, they ask for more self-reflection regarding narratives of democratization.

The report on Quantum and the G77 underlines this argument, highlighting that access to quantum computing is still restricted to a small part of the world, globally speaking [12].

### C. Accessible and diverse education is the pillar for a future quantum workforce

The third broad theme that can be identified in the literature is the relationship between diverse education and workforce building. The World Economic Forum's 2024 Quantum Economy Blueprint emphasizes that a diverse audience must be made aware of the opportunities quantum computing offers in order to attract them as potential future workforce [13]. This is also linked to the theme of avoiding a digital divide. As one of the contributors states, regarding a program called Quantum Leap Africa: "One of the main components of QLA is devoted to quantum training and research, related to and anticipating powerful new [quantum technology], including quantum computing, quantum communication and quantum sensors, to develop experts in Africa who are ready to utilize the power of quantum technology for the benefit of Africa and humanity at large." (Prince Koree Osei, Quantum Leap Africa, Rwanda and Centre President, AIMS Ghana) [13, pp.35]. The WEF report highlights that wide-spread education is a focus in most existing national strategies. It argues that in order to ensure that the benefits of quantum can be reaped across industry, it is important to educate a diverse community to be familiar with quantum computers and how to include them in their respective workflows.

Bridging gaps, emphasizing the need for accessible education, and reflecting on the relationship between education and workforce are the main themes highlighted in the literature focusing on responsible innovation and quantum computing thus far. Taking these themes as backdrop, in the next section of this paper, we will look at some of the challenges in building a diverse community.

## III. STRATEGIES OF BUILDING A DIVERSE COMMUNITY AND HOW TO OVERCOME CHALLENGES

Building and cultivating a diverse community within quantum computing is crucial to ensuring the responsible and equitable development of this transformative technology. The following section discusses strategies implemented by IBM and other stakeholders in an effort to cultivate such a community, and explores the multifaceted challenges encountered in this endeavor. The focus will be on the educational, organizational, and societal barriers to diversity and the innovative solutions implemented to overcome them. While this section primarily focuses on IBM initiatives, we acknowledge that other organizations and institutions are also making significant contributions to fostering diversity in quantum computing. A comprehensive comparison of these efforts across the industry could be a valuable area for future research.

## A. Educational Barriers and Solutions

Quantum computing education often requires a robust foundation in physics, mathematics, and computer science, disciplines that still see underrepresentation of minoritized people. This foundational requirement creates a high entry barrier, limiting the pool of potential quantum professionals. Furthermore, the scarcity of quantum computing programs and resources in underrepresented regions exacerbates this issue [1].

IBM has taken steps to address these barriers through initiatives like the Qiskit Global Summer School and the IBM Quantum Challenge, which provide intensive, hands-on learning experiences aimed at demystifying quantum computing and lowering the entry barriers for a broader audience. The IBM Quantum Learning platform offers free access to a wide range of educational materials, including tutorials and online courses, with the aim of providing comprehensive learning opportunities that are openly and publicly available. Moreover, partnerships with universities, particularly those serving underrepresented communities, such as Historically Black Colleges and Universities (HBCUs), facilitate targeted outreach and support [14].

A notable example of successful educational outreach was the Qubit by Qubit Quantum Course (QXQ). This initiative, a collaboration between IBM Quantum and The Coding School, focused on creating internship pipelines and partnerships with educational institutions to attract diverse talent. The course, which trained over 15,000 students, offered a virtual learning experience covering topics from superposition and entanglement to coding using Qiskit on IBM quantum computers. By offering full and partial scholarships to students and educators with financial need or from underrepresented backgrounds, QXQ ensured that the course was accessible to a diverse cohort of learners. Alumni testimonials include a participant who transitioned from a student to a teaching assistant, and a participant who founded a quantum information science student organization at their own university, exemplifying the program's impact on building a diverse quantum community [15].

Another addition to these efforts in the Asia South region is IBM's sponsorship of the International Collegiate Programming Contest (ICPC) Algo Queen Girls' Programming Cup. This virtual coding competition, organized by Amrita Vishwa Vidyapeetham, aimed to encourage female students from grades 8 to 12 in India to participate in coding competitions. By providing training in competitive coding and quantum computing, this initiative sought to overcome the obstacles that deter girls from pursuing Science, Technology, Engineering, and Mathematics (STEM) fields and promote gender diversity in technology in their region [16].

## B. Organizational Challenges and Solutions

The underrepresentation of women and people from minoritized backgrounds in STEM fields extends into the workplace, where implicit biases and inclusivity challenges can hinder the progress and retention of diverse talent. Additionally, resource disparities, such as limited funding for diversity initiatives and restricted access to quantum computing technology, further impede efforts to build an inclusive quantum community [17].

IBM's approach includes implementing hiring practices intended to promote inclusivity and diversity in recruitment and selection processes. This strategy involves efforts to actively seek candidates from diverse backgrounds and attempts to minimize bias in the evaluation process. Comprehensive diversity and inclusion training programs for all employees create an inclusive workplace culture by educating staff on recognizing and mitigating implicit biases and fostering an environment of respect and collaboration [18].

The Qiskit Advocate program provides individuals who actively contribute to the Qiskit community with opportunities to network with enthusiasts and experts, gain access to Qiskit core members and projects, and attend global events designed for the quantum computing community [19]. The Qiskit Advocate Mentorship Program (QAMP) [20] is an extension of this initiative, offering skill development and professional growth opportunities through hands-on projects guided by IBM Quantum and Qiskit experts. This program helps advocates develop expertise in application areas such as quantum machine learning, quantum chemistry, and quantum finance, and provides professional skills development, fostering a supportive environment for underrepresented groups to thrive. The Qiskit open-source community, guided by a code of conduct, aims to foster a respectful and inclusive environment. This code sets expectations for behavior with the goal of helping community members feel welcome and valued. This code of conduct emphasizes the importance of creating a harassment-free experience and promotes respectful interaction across all community engagements, both online and offline. [21].

Internship programs in quantum computing can play a role in building a diverse pipeline of future professionals. For instance, IBM integrates its efforts through its Quantum Internship Program, which has trained over 400 interns since 2020. These internships provide students with hands-on experience in quantum computing, working directly with researchers and developers. This initiative not only builds technical skills but also cultivates a diverse pipeline of future quantum professionals [22].

## C. Societal Barriers and Solutions

Societal misconceptions about quantum computing as an esoteric and elite field deter wider participation, particularly from marginalized communities. Moreover, limited networking opportunities and exclusive professional networks create challenges for newcomers to integrate and receive support [16].

IBM has initiated outreach and awareness campaigns, including those targeting underrepresented groups, in an effort to address these societal barriers. These campaigns aim to increase understanding of quantum computing and highlight potential avenues for engagement with the field. Virtual events and conferences with low entry barriers, such as the Qiskit Global Summer School, the IBM Quantum Challenge, and other Qiskit events, aim to ensure maximum participation from a global audience, providing a platform for knowledge exchange and networking irrespective of geographical location [23].

The Qiskit open-source community plays a pivotal role in overcoming societal barriers. As an open-source software development kit, Qiskit is allows anyone to contribute to and benefit from the quantum computing community [24]. The Qiskit code of conduct promotes a respectful and inclusive environment, setting clear expectations for behavior and seeking to ensure that all community members feel welcome and valued. Platforms like the Qiskit Slack channel facilitate open dialogue and collaboration among members from varied backgrounds, establishing and nurturing diverse community groups within the quantum computing ecosystem. These initiatives aim to foster a sense of belonging and support, which may play a role in helping newcomers integrate into the field and access resources that could support their growth in quantum computing. [25].

IBM also undertakes efforts to localize Qiskit content for a global community. Based on a survey [26] of 8,709 global consumers in 29 countries in Europe, Asia, North America, and South America, 76% of online users prefer to buy products with information in their native language. The same study also showed that even among people with high proficiency in English, 65% prefer content in their native language. This preference has a major impact on how users consume knowledge and the pace that they can learn a complex topic such as quantum computing.

The Qiskit open-source community has been actively growing a global community by making tutorials, documentation, and event materials available in different languages where the language divide is known to be more prominent than others. Active users of Qiskit around the world have served as volunteer translators [27] making learning content more accessible since October, 2019. User data shows 5 times more traffic to translated material from users who have their browser set in that specific language compared to the traffic from that same cohort to the English original text. Making content and educational material accessible to non-English speaking communities should be regarded an important effort to make sure we are building a diverse global community.

Open access to quantum systems has been a significant factor in democratizing quantum computing. Since 2016, IBM Quantum Experience has enabled public access to quantum computing services via the cloud, allowing anyone with a computer browser and email access to experiment with quantum computing. This open-access approach has made quantum computing more inclusive by lowering the barriers to entry and providing hands-on experience to a broad audience [28]. It is important to note that many quantum computing services, including IBM's, are still unavailable in some countries. [29, p.4, sec. 7]

Platforms that enable user interaction with quantum computers can contribute to increasing understanding of the technology and potentially encourage broader participation in the field. For instance, in 2016, IBM Quantum Experience, the first Quantum Cloud service, opened its doors to the public, allowing users to run quantum circuits on real quantum hardware. This access strategy was instrumental in building a diverse community of enthusiasts and researchers, plugging into the growing quantum ecosystem. By providing a platform where users could interact with actual quantum computers, IBM helped to demystify the technology and encourage widespread participation [30].

## IV. CASE STUDIES INCORPORATED WITHIN BARRIERS AND STRATEGIES

In the context of educational barriers, the Qiskit Global Summer School, The IBM Quantum Challenge and the Qubit by Qubit Quantum Course (QXQ) serve as prime examples of how IBM addresses the challenge of accessibility. The Qiskit Global Summer School and IBM Quantum Challenge provides free, high-quality education to a global audience, lowering entry barriers. The QXQ initiative, with its comprehensive curriculum and substantial scholarships, ensures that students from diverse backgrounds can participate and excel in quantum computing [15].

Organizationally, community-building initiatives such as advocate programs can provide contributors with networking opportunities deeper access to networks such as core development teams. These programs aim to foster engagement and knowledge sharing within the quantum computing community. One such example is the Qiskit Advocate program, which exemplifies IBM's efforts towards fostering an inclusive community. By connecting newcomers with experienced professionals, the program helps mitigate implicit biases and creates a supportive environment. The visibility and recognition of diverse contributions through awards and media representation further support this goal [19].

On a societal level, outreach initiatives like the Qiskit Global Summer School and the QXQ's partnerships with high schools demonstrate effective strategies to engage underrepresented groups. By offering virtual events and comprehensive support, these programs break down geographical and societal barriers, ensuring a diverse and inclusive quantum community [16].

## V. FUTURE WORK

The number of people working on quantum technologies is increasing globally, in part driven by some of the specific efforts and best practices highlighted in this paper. However, more efforts are needed to help decision-makers build a culture of responsibility in quantum technologies and to foster an equitable and inclusive quantum ecosystem.

In one call for proposal for Accelerated Research in Quantum Computing funded by the Department of Energy [31], diversity of supported investigators and institutions and participation of institutions historically underrepresented in the research portfolio was factored into the selection criteria. This is an example of how the public sector can influence decision makers to make more equitable choices and foster a culture of responsible quantum computing. If such practices can become a standard, this can further promote equity and inclusion as an intrinsic element to advancing scientific excellence in quantum research and in building the future industry based on quantum technologies.

As we have seen in Ten Holter [9], responsible innovation requires looking ahead and anticipating future challenges and demands. In doing so, we must ensure that the future workforce focuses beyond the industry today. While we see mainly researchers, academics, and scientists as the major users of quantum computing today, the user base will soon include businesses and industries as the computing capabilities scale and applications became more practical. The future workforce will be further diversified to encompass virtually all industries, government sectors, and individual consumers. Thus, focusing on diversity also in terms of the disciplines that are educated in quantum, we can ensure that the benefits of quantum can be reaped across different sectors.

As we have seen in Seskir et al. [11], democratizing quantum computing is an ongoing challenge. In order to achieve more truly democratic initiatives, collaboration between a broad stakeholder base is necessary. Thus, as this transformative technology continues to evolve, it is paramount that the various stakeholders—including governments, corporations, educational institutions, and individual contributors—collaborate to anticipate and address the potential societal impacts. This collaboration must incorporate a variety of viewpoints, particularly from those most likely to be impacted, to ensure that the benefits of quantum computing are equitably distributed.

## VI. Conclusion

In this paper, we demonstrated that one of the pillars of responsible quantum computing is the incorporation of diverse and inclusive education mechanisms. While this paper primarily focuses on IBM's initiatives, we acknowledge that other organizations and institutions are also making significant contributions to fostering diversity in quantum computing. A comprehensive comparison of these efforts across the industry could be a valuable area for future research.

We also explored the motivations for fostering diversity, defined what diversity entails within the responsible quantum computing framework, described our efforts to build a diverse community, and discussed the challenges encountered.

The journey toward a truly diverse quantum community is fraught with challenges. We identified and explored educational barriers, organizational biases, and societal misconceptions as significant obstacles in this work. Addressing these obstacles requires sustained efforts, innovative solutions, and robust support systems. The examples provided, such as the IBM Internship Program, Qiskit Advocate Program, and Qiskit localization project, along with strategies like open access Quantum Challenges, Qiskit Global Summer School, Qubit by Qubit Quantum Course, and ICPC Algoqueen, inclusive hiring practices, and supportive community initiatives, demonstrate practical steps towards achieving a diverse and inclusive quantum ecosystem. These efforts may offer insights into potential approaches for creating a supportive environment for underrepresented groups, addressing diversity challenges in the field. While the initiatives described have shown promising initial results, it is crucial to continue evaluating and refining these approaches. Ongoing open challenges include ensuring long-term engagement, reaching truly diverse global audiences, and translating educational outreach into increased diversity in the quantum workforce.

As we move forward, continuous evaluation and adaptation of these strategies will be necessary to meet the evolving challenges and opportunities in the field. The ongoing commitment to diversity and inclusion will be essential in helping to maximize the positive impacts of quantum and shaping a quantum future that benefits all of humanity.